\documentclass[conference]{IEEEtran}
\IEEEoverridecommandlockouts
\usepackage{cite}
\usepackage{amsmath,amssymb,amsfonts,amsthm}
\usepackage{algorithmic}
\usepackage{graphicx}
\usepackage{textcomp}
\usepackage{xcolor}
\usepackage{wasysym}
\usepackage{times, epsfig, array, mathrsfs,multirow}
\usepackage[caption=false,font=normalsize,labelfont=sf,textfont=sf]{subfig}
\usepackage{stfloats}
\usepackage{threeparttable}
\usepackage{cases}
\usepackage[linesnumbered, ruled]{algorithm2e}
\SetKwRepeat{Do}{do}{while}%
\usepackage{cite}
\usepackage[all=normal, paragraphs=tight, floats=tight, mathspacing=tight]{savetrees}

\usepackage{comment} 
\newtheorem{prob}{Problem}
\usepackage{bm}

\begin{document}
\title{Interference Mitigation in LEO Constellations with Limited Radio Environment Information
}
\vspace{-0.5cm}
\author{ 
\IEEEauthorblockN{
Fernando Moya Caceres\IEEEauthorrefmark{1},
Akram Al-Hourani\IEEEauthorrefmark{1}
Saman Atapattu\IEEEauthorrefmark{1},
Michael Aygur\IEEEauthorrefmark{1},\\
Sithamparanathan Kandeepan\IEEEauthorrefmark{1},
Jing Fu\IEEEauthorrefmark{1},
Ke Wang\IEEEauthorrefmark{1}, 
Wayne S. T. Rowe\IEEEauthorrefmark{1},
Mark Bowyer\IEEEauthorrefmark{2},\\
Zarko Krusevac\IEEEauthorrefmark{3},
and Edward Arbon\IEEEauthorrefmark{3}}
 \IEEEauthorblockA{
 \IEEEauthorrefmark{1}School of Engineering, RMIT University, Melbourne, Victoria, Australia. \\
 \IEEEauthorrefmark{2}Airbus Defence and Space, Portsmouth, United Kingdom. \\
 \IEEEauthorrefmark{3}The Defence, Science and Technology Group, Canberra, Australia. \\
\IEEEauthorblockA{Email:
\IEEEauthorrefmark{1}\{fernando.moyacaceres, akram.hourani\}@rmit.edu.au;\,
}
}
\vspace{-1.0cm}
}

\maketitle

\begin{abstract}
This research paper delves into interference mitigation within Low Earth Orbit (LEO) satellite constellations, particularly when operating under constraints of limited radio environment information. Leveraging cognitive capabilities facilitated by the Radio Environment Map (REM), we explore strategies to mitigate the impact of both intentional and unintentional interference using planar antenna array (PAA) beamforming techniques. We address the complexities encountered in the design of beamforming weights, a challenge exacerbated by the array size and the increasing number of directions of interest and avoidance. Furthermore, we conduct an extensive analysis of beamforming performance from various perspectives associated with limited REM information: static versus dynamic, partial versus full, and perfect versus imperfect. To substantiate our findings, we provide simulation results and offer conclusions based on the outcomes of our investigation.    
\end{abstract}

\begin{IEEEkeywords}
Beamforming, Interference mitigation, Low Earth Orbit (LEO) satellite, Planar antenna array, Radio Environment Map (REM)
\end{IEEEkeywords}

\section{Introduction} 
The increasing congestion of the frequency spectrum is a well-recognized issue impacting both terrestrial and satellite applications. This congestion primarily stems from evolving user demands in established and emerging technologies~\cite{ITU-Mag2023}. The International Telecommunication Union (ITU) consistently publishes new documents, with a notable focus on the 5G spectrum in 2019~\cite{ITU-Mag2019} and, as of 2023, a strong emphasis on space services~\cite{ITU-Mag2023}. One particularly noteworthy aspect of space services relates to Satellite Communication (SatCom) systems, which have attracted significant attention from entities exploring non-geostationary satellite orbits (non-GSO) constellations~\cite{ITU-Mag2023}. Technological advancements in non-GSO constellations hold the promise of delivering high throughput with minimal latency, representing a substantial improvement over traditional SatCom systems~\cite{9210567}.

Prominent international non-geostationary satellite orbit (non-GSO) constellations operating in Low Earth Orbit (LEO) and providing broadband Internet services include SpaceX's Starlink, with a grand total of 41,927 planned satellites, Amazon's Project Kuiper, with plans to deploy 3,236 satellites, and OneWeb's OneWeb, which envisions a constellation of 1,980 satellites~\cite{10050615}. These extensive non-GSO constellations, offering Internet connectivity to users via very small aperture terminals (VSAT), are commonly referred to as ``mega-constellations" by both the ITU and various sources in the literature~\cite{ITU-Mag2019}. A critical factor enabling the use of compact antennas is the choice of frequency bands. Notably, Ka-, Ku-, Q-, and V-bands have been adopted for SatComs, as highlighted in publications by the ITU in both 2019 and 2023~\cite{ITU-Mag2019,ITU-Mag2023}.

Technological progress in this field grapples with radio frequency interference (RFI) stemming from both deliberate sources (e.g., jammers) and unintentional sources within mega-constellations~\cite{9755278}. Deliberate interference seeks to disrupt access through various jamming methods, such as constant, reactive, or random/periodic jamming~\cite{9733393}, and the deployment of smart jammers~\cite{10048559}. Jamming, without effective mitigation, can disrupt communication infrastructures and degrade system performance. Interference mitigation strategies encompass temporal, spectral, power, and spatial domains, aiming to safeguard the host system to varying degrees. In SatCom systems, the use of antenna arrays introduces spatial degrees of freedom (DoF)\cite{7762075}. One spatial mitigation technique is beamforming, enhancing physical layer security (PLS) by creating nulls in the antenna radiation pattern directed at interference sources and shaping the pattern to improve communication for legitimate users\cite{7762075,9402861,8701493,9072611}.

The adoption of higher frequency bands has resulted in significantly smaller satellite antennas, enabling the use of denser on-board antenna arrays, which are more effective in mitigating interference. Numerous methods for optimizing beamforming weights are available~\cite{Balanis-2012-antenna}. However, these methods become analytically complex when dealing with multiple arrival directions of interest and avoidance concurrently. Additionally, as the number of antennas grows, the design complexity escalates, making it increasingly challenging to efficiently optimize the array weights. In this study, we explore interference mitigation in LEO constellations using beamforming, even in scenarios with limited information about the radio environment. Our contributions are as follows:
\begin{enumerate}
    \item A comprehensive evaluation and comparison of beamforming optimization algorithms, namely Intermediate Point (IP), Global Search (GS), Genetic Algorithm (GA), Pattern Search (PS), Particle Swarm (PSW), and Simulated Annealing Algorithm (SAA), for interference mitigation in LEO scenarios.
    \item An in-depth study of interference mitigation performance considering three distinct Radio Environment Map (REM) information perspectives: static versus dynamic, partial versus full, and perfect versus imperfect.
\end{enumerate}


\section{System Model}\label{Sec_SysModel}
\begin{figure}
    \centering
    \includegraphics[width=0.45\textwidth]{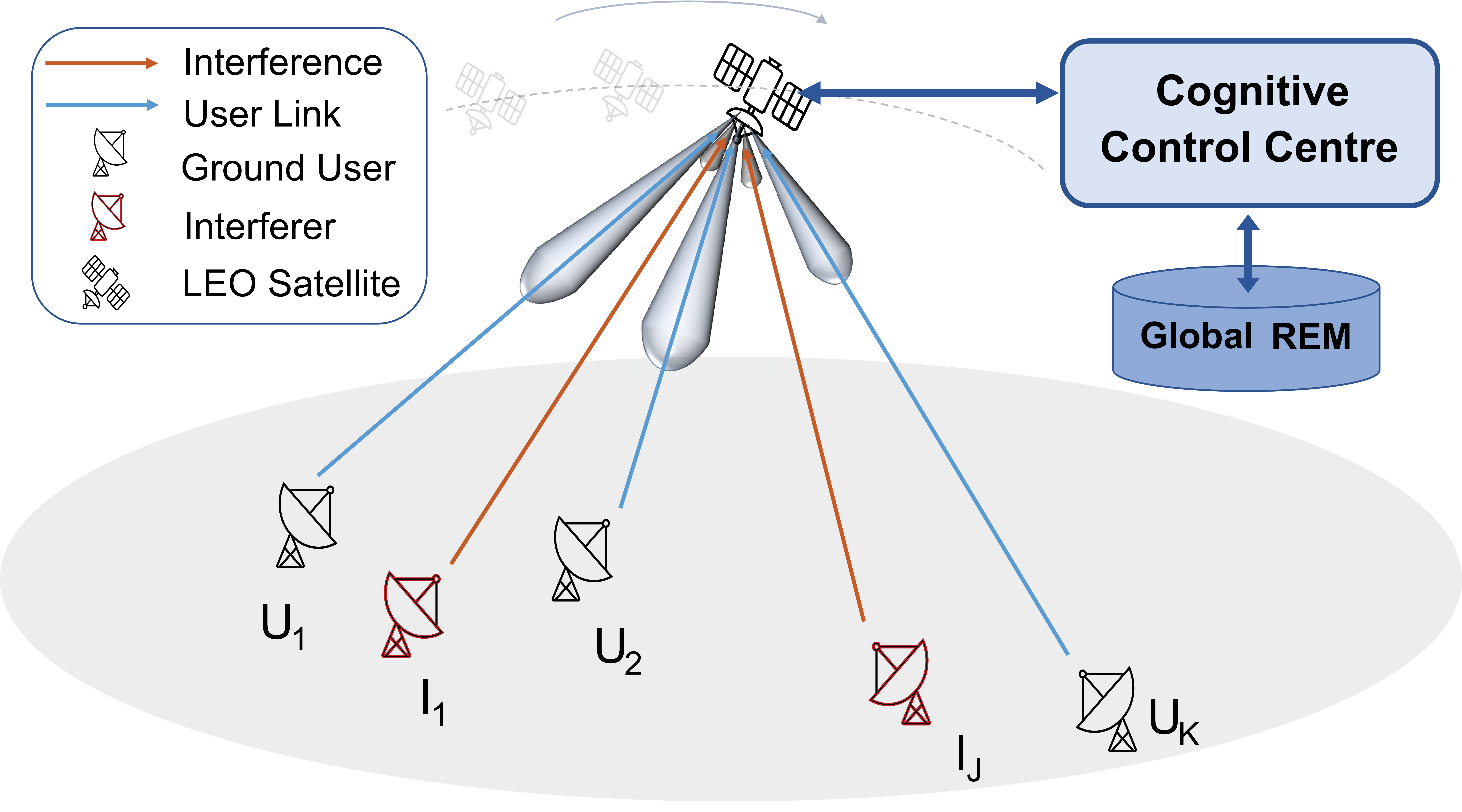}
    \caption{Cognitive LEO ground to satellite (uplink) in the presence of interference.}
    \label{fig:Sys_Mod}
\end{figure}
In our scenario, we have a Cognitive SatCom LEO system serving \(K\) ground users, within the coverage spot there are also \(J\) interference sources affecting the uplink signal, as illustrated in Fig.~\ref{fig:Sys_Mod}. The REM available to the satellite stores spectrum occupancy data, encompassing details like geo-location, carrier frequencies, and transmission power levels, thus supporting cognitive functions. Additionally, the satellite employs beamforming capabilities to counter interference, utilizing information received from the REM.

\subsection{REM Model}
The REM enhances SatCom network awareness by estimating location information about interfering sources. This data is obtained through both direct observation and network assistance provided to the REM. For this study, we presume that relevant spectrum occupancy information is accessible to the REM with various levels of accuracy and completeness. As depicted in Fig.~\ref{fig:Sys_Mod}, the REM receives queries from the Satellite Cognitive Control Center (CCC), which oversees the cognitive control plane. The CCC could be on the satellite itself or de-centrlized among the constellation. 

\subsection{Beamforming Model}
\begin{figure}
     \centering
         \includegraphics[width=0.43\textwidth]{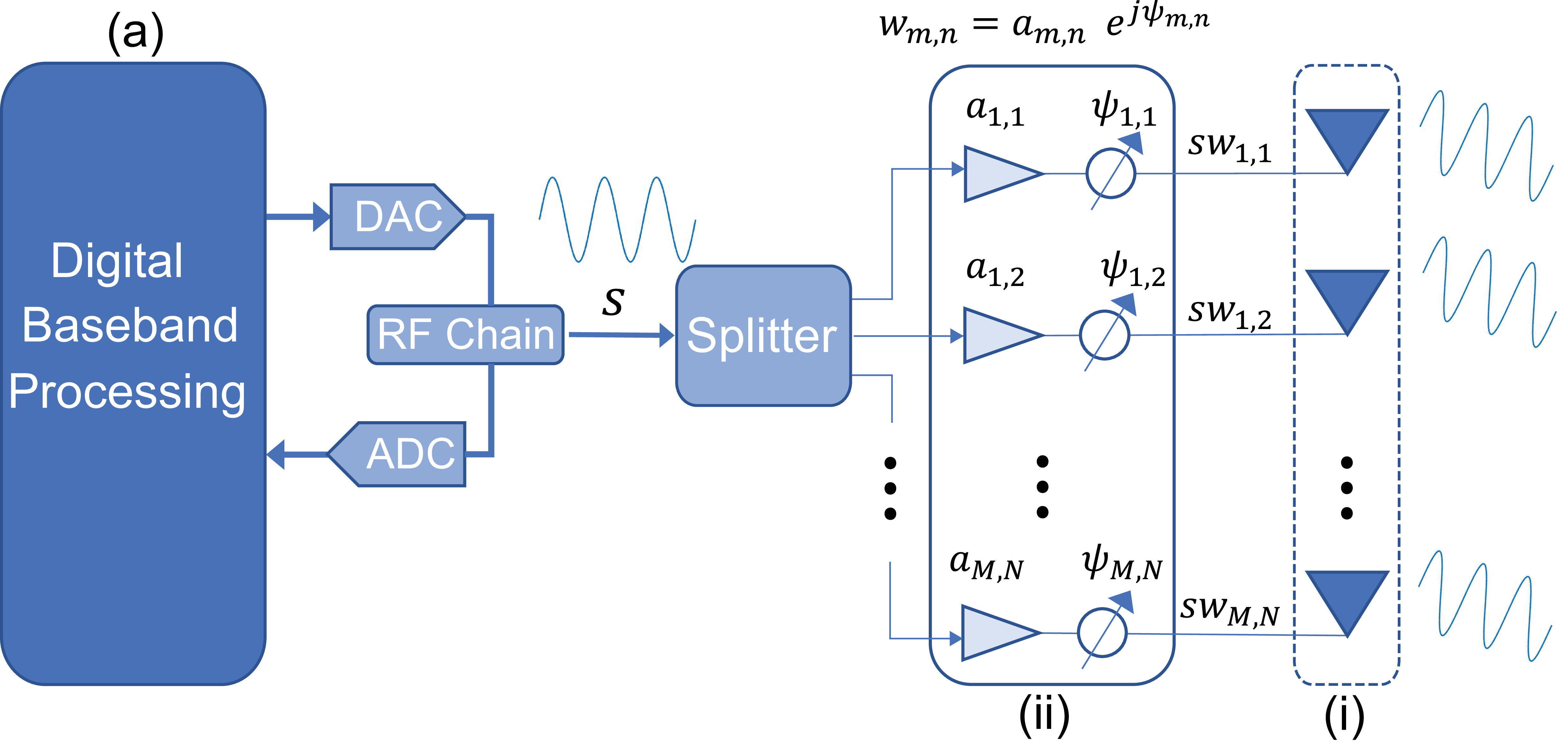}
           \includegraphics[width=0.43\textwidth]{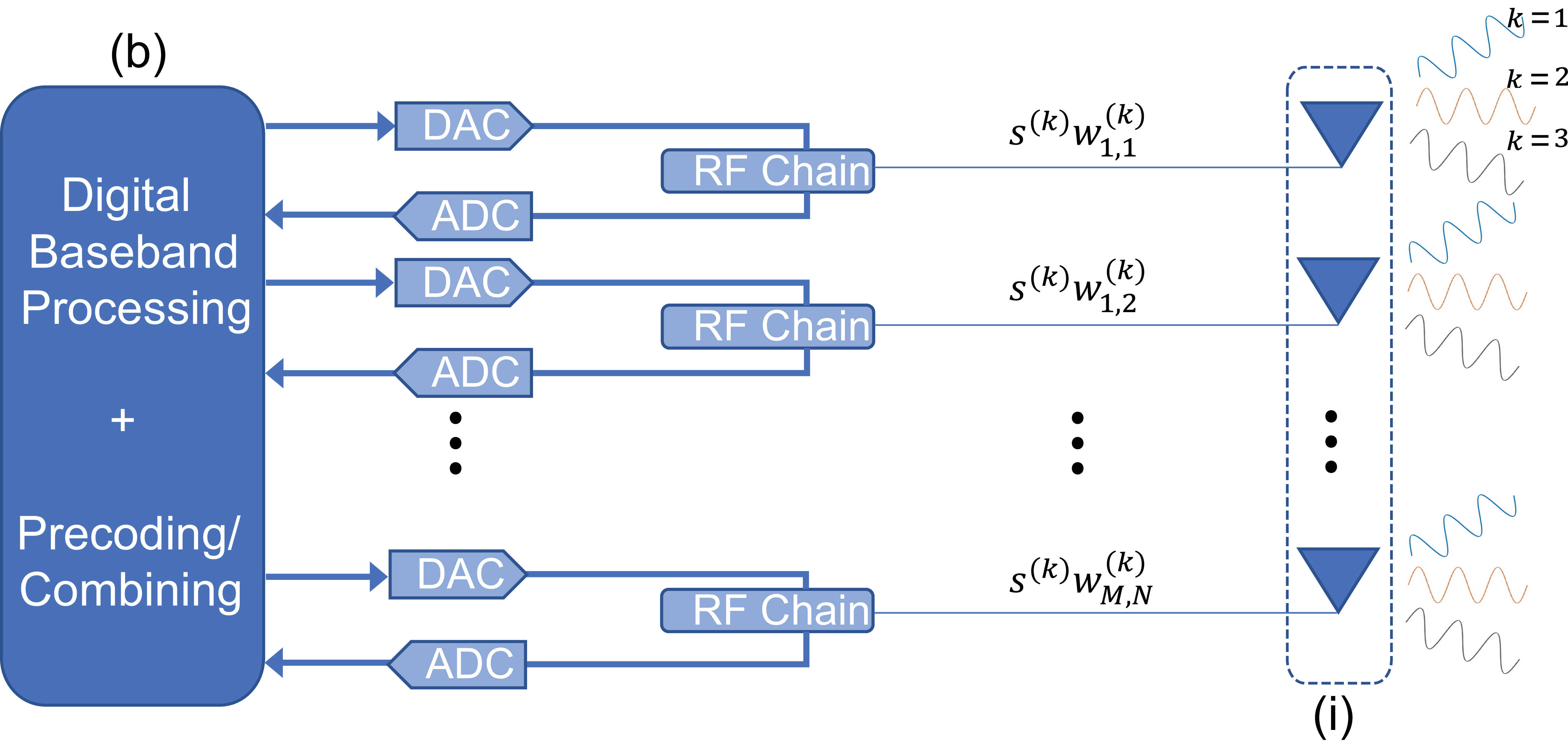}
          \caption{Beamforming architectures: (a) analog beamforming with a single datastream, (b) fully digital beamforming with multiple datastreams.}
    \label{fig:BF_Options}
\end{figure}
\begin{figure}[!ht]
\centering
\includegraphics[width=0.43\textwidth]{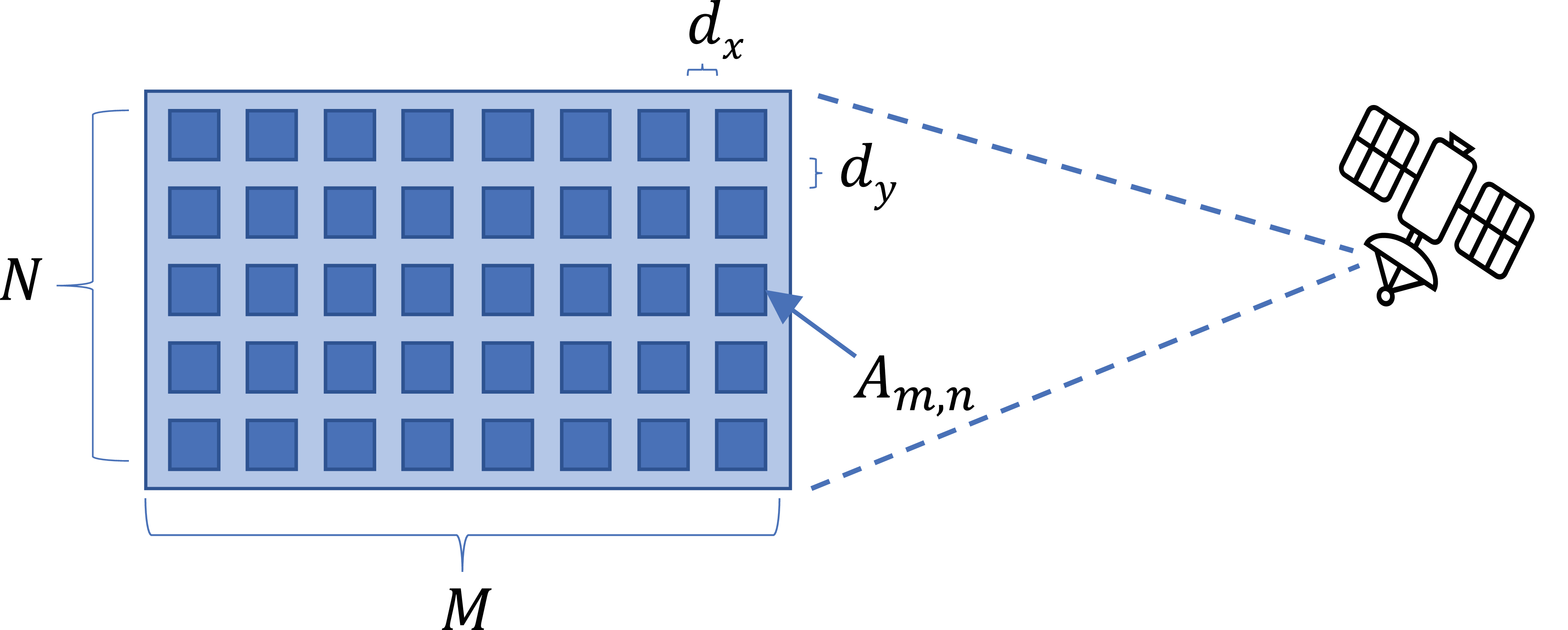}
\caption{Satellite \(M\times N\) antenna array.}
\vspace{-5mm}
\label{fig:Ant_Array}
\end{figure}
Beamforming is executed in analog, digital, or hybrid mode, as detailed in Fig.~\ref{fig:BF_Options}:
\begin{enumerate}
    \item Analog beamforming involves the sequential transmission of a single data stream. As depicted in Fig.~\ref{fig:BF_Options}-(a), the signal is divided and directed to analog phase shifters (ii), where corresponding weights are applied. It is then transmitted via an antenna array (i).
    \item Digital beamforming, as in Fig.~\ref{fig:BF_Options}-(b), handles multiple data streams concurrently with digital weighting factors customized for each user. The final step involves transmitting these signals through the antenna array (i). 
    \item Hybrid beamforming is a combination of analog and digital beamforming balancing complexity and performance. 
\end{enumerate}

We assume that the satellite is equipped with an \(M \times N\) planar antenna array, as depicted in Fig.~\ref{fig:Ant_Array}. This array is capable of dynamically generating multiple beams to enhance system performance and counteract interference effects. Each antenna element \(A_{m,n}(a, \psi)\), with \(m \in M\) and \(n \in N\), is associated with a unique excitation amplitude \(a_{m,n}\) and phase \(\psi_{m,n}\), collectively forming a weighting factor \(w_{m,n} = a_{m,n}e^{j\psi_{m,n}}\). The distances between the elements are defined by \(d_x\) and \(d_y\) along the \(X\) and \(Y\) axis respectively. For a given arrival direction described by the angles \(\theta\) and \(\phi\), the total electric field radiated (or received) by the antenna array is given as~\cite{Balanis-2012-antenna},
\begin{equation}
\label{ant_rad}
    E_\mathrm{Tot}(\phi,\theta)=\mathrm{EF}(\phi,\theta)\times \mathrm{AF}(\phi,\theta)
\end{equation}
where \(\mathrm{EF}(\phi,\theta)\) is the antenna element factor, i.e. gain pattern, and \(\mathrm{AF}(\phi,\theta)\) is the array factor.  In this study, for simplicity, we consider the antenna elements to be isotropic, meaning that \(\mathrm{EF}(\phi,\theta)=1\). Further, the array factor is given as 
\begin{equation}
\label{AF_eq}
    \mathrm{AF}\!(\phi,\theta)\!=\!\!\!\sum_{m\!=\!1}^M\!\!\sum_{n\!=\!1}^N w_{m,n}e^{-j\frac{2\pi}{\lambda}\left((\!m\!-\!1\!)d_x\sin{\theta}\cos{\phi}+(\!n\!-\!1\!)d_y\sin{\theta}\sin{\phi}\right)}.
\end{equation}
The position of the antenna element \(A_{m,n}\) relative to the array's origin is determined by the vector \(\mathbf{r_{m,n}} = \left[x_{m,n}, y_{m,n}, z_{m,n}\right]\). It is important to note that in this context, \(z_{m,n} = 0\). The wave-vector characterizes the phase variation of a plane wave in three orthogonal directions. Its magnitude represents the wave number \(k = \frac{2\pi}{\lambda}\), and can be given as
\begin{equation}
\label{wave_vector}
    \mathbf{k}=(k_x,~k_y,~k_z)=\frac{2\pi}{\lambda}(\sin{\theta}\cos{\phi},~\sin{\theta}\sin{\phi},~\cos{\theta}).
\end{equation}
We also express the steering vector \(\mathbf{V(k)}\) as the vector of propagation delays across the planar array for a given wave-vector \(\mathbf{k}\) as 
\begin{equation}
\label{steer_vec}
\mathbf{V(k)} =
\begin{bmatrix}
e^{-j\mathbf{k}\mathbf{r_{1,1}}} & e^{-j\mathbf{k}\mathbf{r_{1,2}}} & \dotsc &e^{-j\mathbf{k}\mathbf{r_{1,n}}} \\
e^{-j\mathbf{k}\mathbf{r_{2,1}}} & e^{-j\mathbf{k}\mathbf{r_{2,2}}} & \dotsc &e^{-j\mathbf{k}\mathbf{r_{2,n}}} \\
\vdots & \vdots & \ddots &\vdots \\
e^{-j\mathbf{k}\mathbf{r_{m,1}}} & e^{-j\mathbf{k}\mathbf{r_{m,2}}} & \dotsc &e^{-j\mathbf{k}\mathbf{r_{m,n}}} \\
\end{bmatrix}.
\end{equation}
The antenna array output \(Y\) is then written as
\begin{equation}
\label{Ant_arr_out}
    Y=\sum_{m=1}^M\sum_{n=1}^N w_{m,n}X_{m,n}
\end{equation}
where \(X_{m,n}\) represents the input of the antenna element \(A_{m,n}\) within the array. To facilitate the analysis, let us define each of the antenna column inputs in vector format such that \(\mathbf{X_n} = (X_{1,n}, \dotsc, X_{M,n})\), and then consolidate all column vectors into a single column vector \(\mathbf{X} = (X_1, \dotsc, X_N) \in \mathbb{C}^{MN\times1}\). Utilizing a similar logic for the construction of \(\mathbf{W} \in \mathbb{C}^{MN\times1}\), \eqref{Ant_arr_out} can be re-written as 
\begin{equation}
\label{Ant_arr_out_alt}
    Y=\mathbf{W}^T\mathbf{X}.
\end{equation}

\subsection{Geometric Model}
The REM possesses knowledge of the positions of both the ground users and the interfering sources using Earth-centered Earth-fixed (ECEF) Cartesian coordinates \((X, Y, Z)\) system, which has the Earth at its center and rotates along with the Earth. To calculate the direction of arrival angles, the satellite converts these coordinates to its local North-East-Down (NED) frame. These NED coordinates are then transformed from Cartesian to spherical coordinates, yielding the position of each point in terms of azimuth \(\theta\), off-nadir angle \(\phi\), and range \(\rho\). Consequently, \(U_k = (\theta_k, \phi_k, \rho_k)\) and \(I_j = (\theta_j, \phi_j, \rho_j)\) represent the locations of user \(U_k\) and interferer \(I_j\) relative to the satellite's position.

\begin{figure}
\centering
\includegraphics[width=0.45\textwidth]{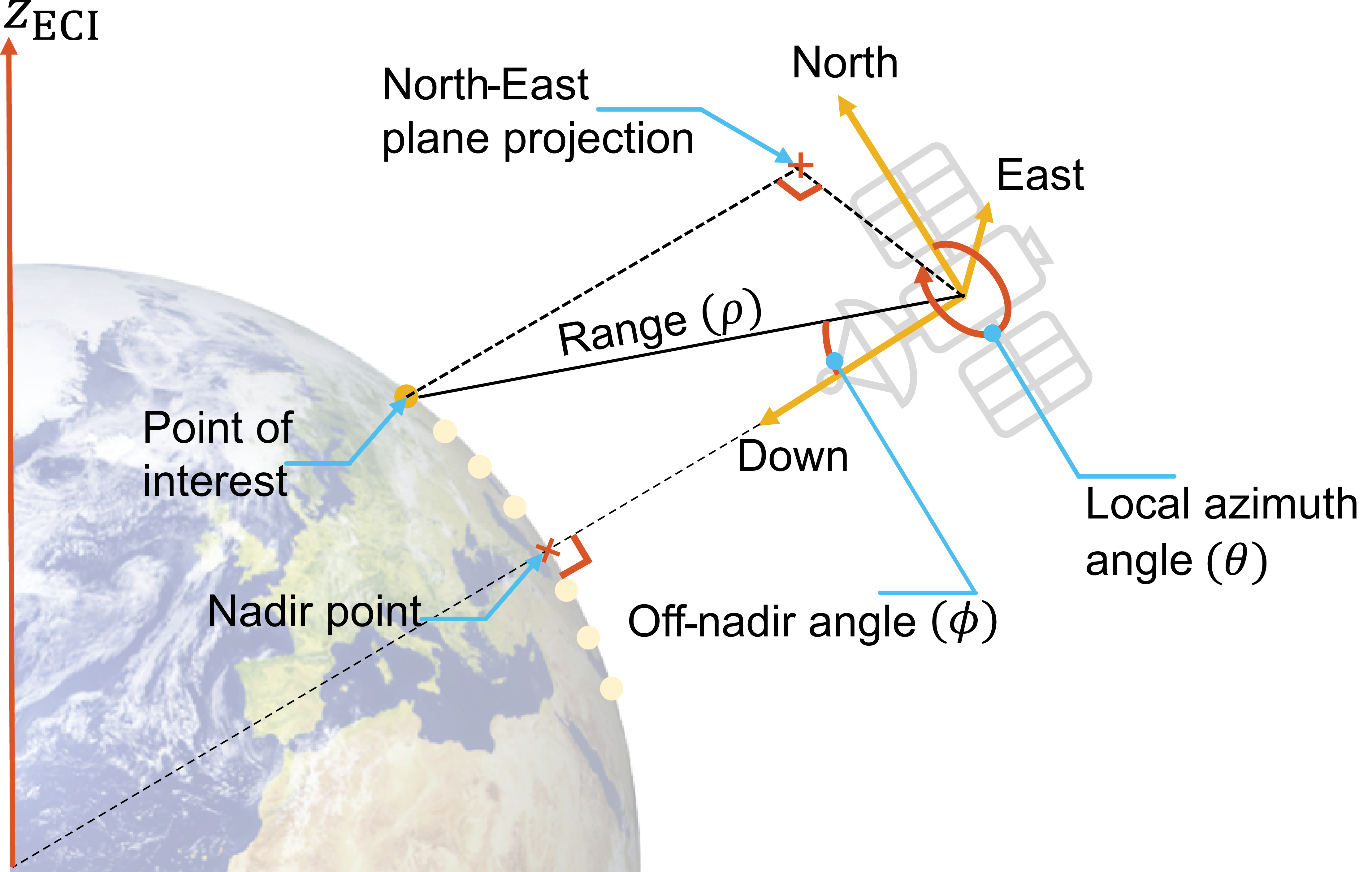}
\caption{NED frame of reference centered on the LEO satellite.}
\label{fig:Geometric_Model}
\end{figure}

\subsection{Channel Model}
We define \(\mathbf{H}_k\) as the channel vector between the satellite and ground station \(k\) (whether legitimate or interfering), expressed as
\begin{equation}
\label{eq_channel}
\mathbf{H}_k = L\mathbf{V(k)},
\end{equation}
where for the purposes of this study, we account for the effects of free-space path loss (FSPL),  defined as 
\begin{equation}
\label{eq_PL}
L = \left(\frac{\lambda}{4\pi\rho_k}\right)^2
\end{equation}
where \(\lambda\) is the transmission wavelength.


\subsection{Signal Model}
In the uplink direction, we assume the presence of \(J\) interferers disrupting the communication link between the ground users and the satellite. The received signal at the satellite from user \(k\) at time instance \(t\) is given as
\begin{equation}
\label{Received_signal_satellite}
y_s\left(t\right)=\mathbf{W}^T\left(\mathbf{H_k}\sqrt{P_k}s_k(t)+\sum_{j=1}^J\mathbf{H_j}\sqrt{P_j}s_j(t) +\mathbf{N}(t)\right)
\end{equation}
where \(P_k\) and \(P_j\) represent the effective transmit power (EIRP) of user \(k\) and interferer \(j\), respectively, \(\mathbf{N}(t)\sim \mathbb{CN}(0,\sigma^2I)\) is the additive Gaussian white noise (AWGN) matrix with zero mean and variance \(\sigma^2=\kappa B T\), with \(\kappa=1.38\times10^{-23}J/K\) denoting Boltzmann constant, \(T\) the noise temperature and \(B\) the noise bandwidth. With respect to the $k^\mathrm{th}$ user signal. Accordingly, the signal-to-interference-and-noise ratio (SINR) at the satellite is given as 
\begin{equation}
\label{SINR_satellite}
    \gamma_{s,k}(t)=\frac{P_k|\mathbf{W}^H\mathbf{H_k}|^2}{|\sum_{j=1}^J P_j\mathbf{W}^H\mathbf{H_j} + \mathbf{W}^H\mathbf{N}(t)|^2}.
\end{equation}
Then, the throughput \(C_{s,k}\) from user \(k\) at the satellite, is 
\begin{equation}
\label{Thr_satellite}
    C_{s,k}(t)=B\times\log_2\left(1+\gamma_{s,k}(t)\right)
\end{equation}
Subsequently, the total capacity of the satellite  system, considering $K$ users, is calculated as 
\begin{equation}
\label{Thr_tot_satellite}
    C_\mathrm{Tot}(t)=\Lambda\sum_{k=1}^K C_{s,k}(t).
\end{equation}
where \(\Lambda\) is \(1/K\) for analog beamforming and $1$ for digital beamforming.

\section{Interference Mitigation Framework}\label{Sec_IMF}
In this section, our goal is to design the receive beamforming mechanism to maximize the total capacity of the satellite system, given the knowledge of interference locations and strengths. We assume that user nodes and interference nodes transmit signals at their maximum powers, denoted as $P^\mathrm{{User}}$ and $P^\mathrm{{Int}}$, respectively. For time $t$, our objective is to find the optimal values for $\mathbf{W}$, specifically $a_{m,n}$ and $\psi_{m,n}$, $\forall m,n$, to maximize the received signal from the desired users while simultaneously introducing minima/nulls towards the interferers. This is equivalent to the following optimization problem,
\begin{prob}\label{p:ee-1-eqv}
\begin{subequations}
\begin{eqnarray}
\mathop{\max} \limits_{\bm{\mathbf{W}}} \quad && C_\mathrm{Tot}(t,a_{m,n}, \psi_{m,n}, \theta_k, \phi_k, \rho_k,\theta_j, \phi_j, \rho_j) \nonumber \\
\text{s.t.} \quad && 0 \le \sum_{m=1}^{M}{\sum_{n=1}^{N}{a_{m,n}}} \le 1  \label{e:ee-1-gamma1}\\
            \quad && 0 \le  \psi_{m,n} \le 2\pi; \quad\forall m\in M, \forall n\in N \label{e:ee-1-gamma2},
\end{eqnarray}
\end{subequations}
\end{prob}
which presents a challenge in finding an analytic solution due to the non-convex nature of the function $C_\mathrm{Tot}(\cdot)$ with respect to $a_{m,n}$ and $\psi_{m,n}$ for all $m\in M$ and $n\in N$. As the number of users and antenna array size increases, the optimization process becomes computationally expensive due to the large number of weights that need to be computed. Since our primary focus is on system-level interference mitigation evaluations under various practical constraints, we do not provide a detailed procedure for solving Problem~\ref{p:ee-1-eqv}. Instead, we employ optimization algorithms available in software packages such as MATLAB.  

\begin{figure}
     \centering
        \includegraphics[width=0.43\textwidth]{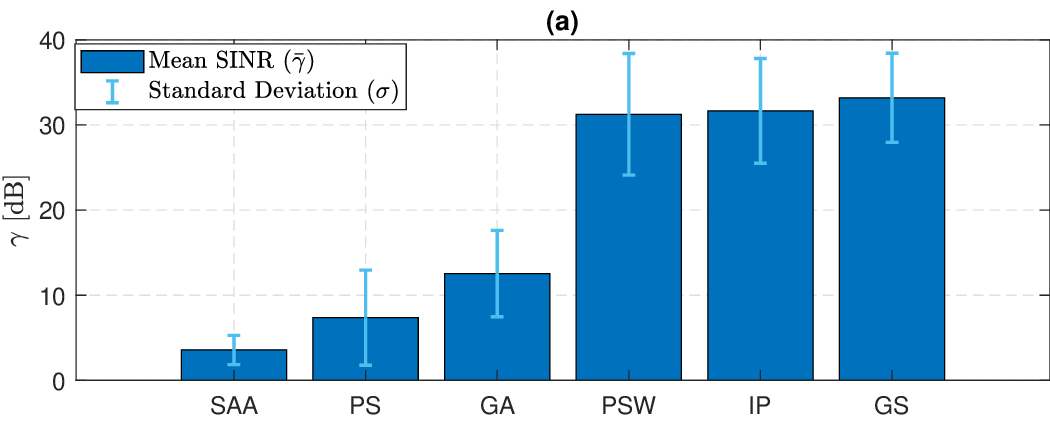} 
         \includegraphics[width=0.43\textwidth]{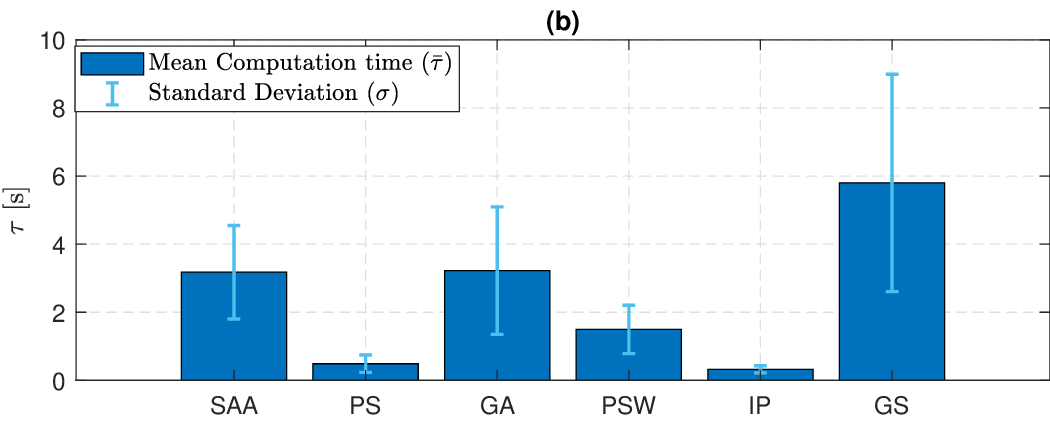}   
         \caption{Optimization algorithms' performance for \(M\times N=16\) beamforming: Simulated Annealing Algorithm (SAA), Pattern Search (PS), Genetic Algorithm (GA), Particle Swarm (PSW), Intermediate Point (IP) and Global Search (GS). (a) SINR \(\gamma\) performance, (b) Computation time \(\tau\).}
    \label{fig:Opt_Algo_Matlab}
\end{figure}

In Fig.~\ref{fig:Opt_Algo_Matlab}, a comparison of various optimization algorithms is presented, assuming a planar array of size \(M\times N=16\) over $1500$ runs. Fig.~\ref{fig:Opt_Algo_Matlab}-(a) illustrates the achievable optimal SINR performance, revealing significant variation in the obtained \(\gamma\) depending on the algorithm used. This variability is attributed to the random initialization process and the stochastic nature of how each algorithm seeks a solution. Notably, PSW, IP, and GS exhibit the best performance. In Fig.~\ref{fig:Opt_Algo_Matlab}-(b), the mean computation time \(\Bar{\tau}\) required to obtain the optimal solution is depicted. Among the algorithms that achieved the best mean SINR, IP stands out as the fastest, with PSW coming in second in terms of computational time. Despite GS showing the best optimization performance, IP emerges as the most balanced choice, offering an optimal 
\(\Bar{\gamma}\) and \(\Bar{\tau}\) compromise.

\begin{figure}
    \centering
    \includegraphics[width=0.43\textwidth]{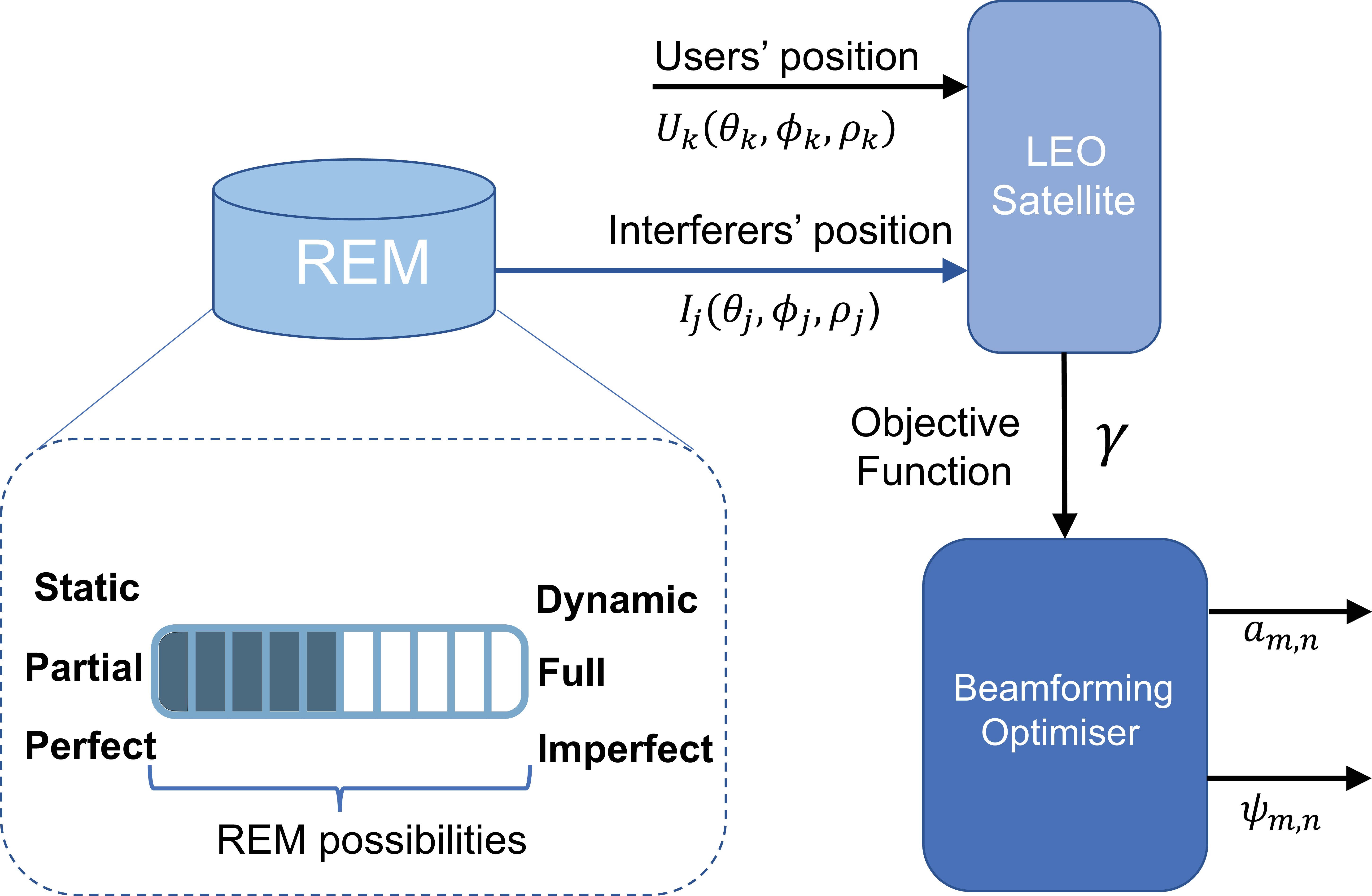}
    \caption{Cognitive beamforming optimisation framework for limited radio environment information.}
    \label{fig:Opt_Model}
\end{figure}

The proposed optimization framework is depicted in Fig.~\ref{fig:Opt_Model}. The effectiveness of interference mitigation greatly depends on the information provided to the LEO satellite by the REM. Hence, the interest is to assess interference mitigation capabilities under various REM information conditions.

\subsubsection{Static vs Dynamic REM}
The dynamism of the REM refers to how frequently it updates its information, providing the latest radio environment data to the network. This update interval, denoted as \(\Delta t\), directly influences the mitigation performance of the satellite system, especially in the case of LEO satellites, which has an inherent fast relative motion. From the perspective of ground stations, users and interferers can be either mobile or static, making it challenging to obtain REM updates. In cases where users and interferers are static, REM information must be updated to account for the deterministic movement of LEO satellites. Consequently, the arrival angles at which the satellite perceives the ground stations change rapidly during a satellite pass~\cite{9422812,rs15235603}. When a new REM update is unavailable, the satellite continues to use beamforming with the last optimal weights calculated from the most recent REM update. Thus, the satellite updates its beamforming weights every \(\Delta t\) seconds. The optimization process occurs at \(t=i\Delta t\), where \(i\) represents a sequential number denoting the current time step update, as described in the following objective function: 
\begin{equation}
\max_{\mathbf{W}}\quad{C_\mathrm{Tot}(i\Delta t,a_{m,n}, \psi_{m,n}, \theta_k, \phi_k, \rho_k,\theta_j}, \phi_j, \rho_j, K, J).\label{eq:opt_obj_ul_REM_delta}
\end{equation}
Then, all bit rate capacity calculations until the next REM update is received, are computed using the optimal weight vector \(\mathbf{W^*}_{t=n\Delta t}\) as 
\begin{equation}
\label{Thr_tot_satellite_opt}
    C_\mathrm{Tot}(t,\mathbf{W^*}_{t=i\Delta t})=\sum_{k=1}^K C_{s,k}(t), \hspace{0.1cm}\forall i \Delta t\leq t<(i\!+\!1) \Delta t.
\end{equation}

\subsubsection{Partial vs Full REM}
The REM may only possess knowledge of a subset of the interferers within the network. In such a scenario, only partial information regarding the actual radio environment is conveyed to the satellite, and the beamforming process relies on this limited input data. If the REM is aware of only \(Q\) interferers, where \(Q<J\), then the objective function is expressed as 
\begin{equation}
\max_{\mathbf{W}}\quad{C_\mathrm{Tot}(t,a_{m,n}, \psi_{m,n}, \theta_k, \phi_k, \rho_k,\theta_j}, \phi_j, \rho_j, K, Q).\label{eq:opt_obj_ul_REM_partial}
\end{equation}
The computation of beamforming weights takes into account the presence of \(Q\) interferers, and the assessment of interference mitigation effectiveness is conducted in the presence of \(J\) interferers affecting the uplink.

\subsubsection{Perfect vs Imperfect REM}
Estimation algorithms used to construct the REM may introduce errors in the cognitive data supplied to the network. Consequently, the accuracy issues affect all decisions within the cognitive network, thereby restricting its overall performance. We model the error in the estimation of arrival angles as
\begin{subequations}
\begin{align}
     \hat{\theta_j}=\theta_j+\theta_\mathrm{{error}}\text{ and } \hat{\phi_j}=\phi_j+\phi_\mathrm{{error}}.\label{Eq:estimated_phi}
\end{align}    
\end{subequations}
The resulting error, measured in kilometers, arises from the disparity between the actual interferer's position and the estimated position. Consequently, the objective function of  optimization Problem~1 is modified as 
\begin{equation}
\max_{\mathbf{W}}\quad{C_\mathrm{Tot}(t,a_{m,n}, \psi_{m,n}, \theta_k, \phi_k, \rho_k,\hat{\theta_j}, \hat{\phi_j}, \rho_j)}\label{eq:opt_obj_ul_REM_error}
\end{equation}
Finally, \(C_\mathrm{Tot}\)  is computed using the actual values of \(\theta_j\) and \(\phi_j\) to quantify the performance difference attributable to the estimation error.

\section{Numerical results}\label{Sec_Results}
Numerical simulations were conducted employing a LEO satellite positioned at an 800~km altitude, operating in the L-band,  with orbit parameters outlined in Table~\ref{tb:sim_parameters}. 
\vspace{-2mm}
\begin{table}[h!]
\centering
\caption{LEO Satellite Orbital Elements.}
\begin{tabular}{|l|l|}
 \hline
 Variable Description & Value \\
 \hline
 Semi Major Axis, \(a\) & 7173 [km]\\ 
 Eccentricity, \(e\) & \(\approx\) 0\\ 
 Inclination, \(i\) & 86.39\(^{\circ}\)\\
 Right Ascension of Ascending Node, \(\Omega\) & 146.16\(^{\circ}\)\\
 Argument of Periapsis, \(\omega\) & 269.5\(^{\circ}\)\\
 True Anormality, \(v\) & 0.6\\
 \hline
\end{tabular} 
\label{tb:sim_parameters}
\vspace{-2mm}
\end{table}

\begin{figure}
\centering
\includegraphics[width=0.45\textwidth]{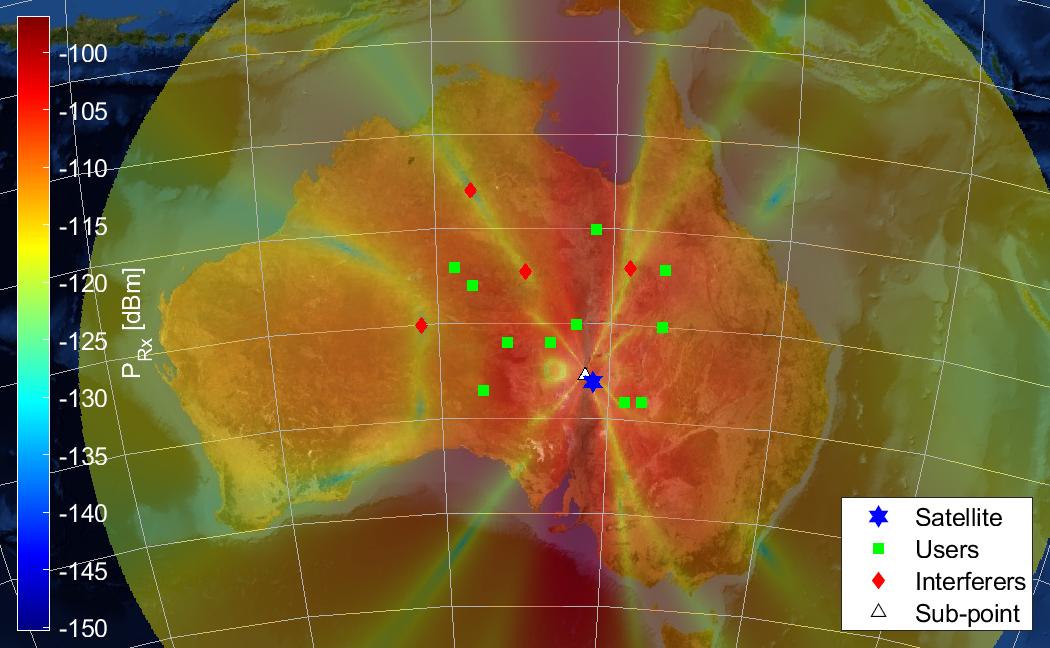}
\caption{Perceived power levels from LEO satellite's perspective over Australia employing interference mitigation with \(8\times 8\) antenna array beamforming under dynamic, perfect, full REM conditions for \(K=11\) and \(J=4\).}
\label{fig:BF_footprint}
\end{figure}

Initially, a dynamic, comprehensive REM scenario was chosen to validate the optimizer's efficacy. Fig.~ \ref{fig:BF_footprint} displays the received power levels at the LEO satellite for transmissions originating from any point across the surface of Australia, employing an \(8\times 8\) antenna array. The green squares denote \(K=11\) fixed ground users, while the red diamonds represent \(J=4\) fixed interferers distributed among the legitimate ground users, causing interference in the uplink. The graph illustrates the creation of nulls around the interferers' positions to mitigate the effects of transmissions to the satellite from those locations. Simultaneously, the radiation pattern of the antenna array is shaped and directed towards the positions of the users to maximize the received power levels from their coordinates at the satellite.
\begin{figure}
\centering
\includegraphics[width=\linewidth]{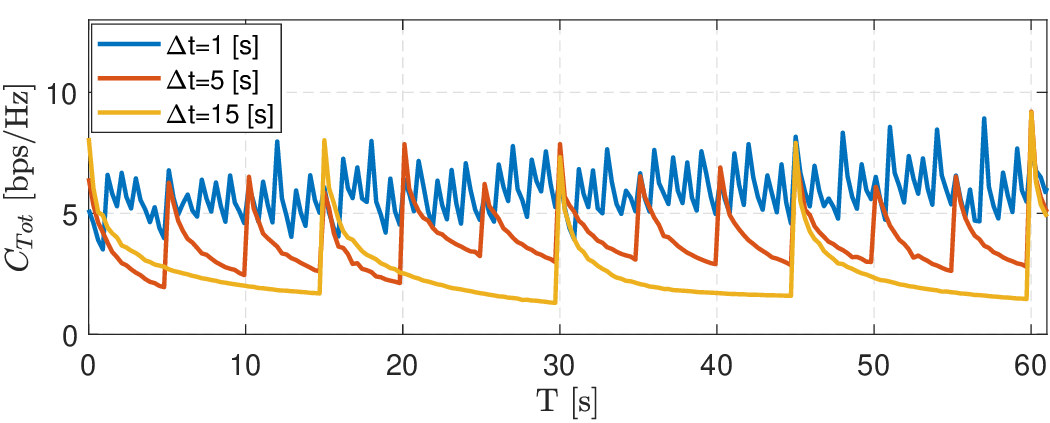}
\caption{Analog beamforming \(M\times N=16\): Total system capacity performance with dynamic REM vs LEO orbiting time for \(K=2\) and \(J=2\).}
\label{fig:Analog_BF_Frequently_Updated_REM}
\end{figure}
\begin{figure}
\centering
\includegraphics[width=\linewidth]{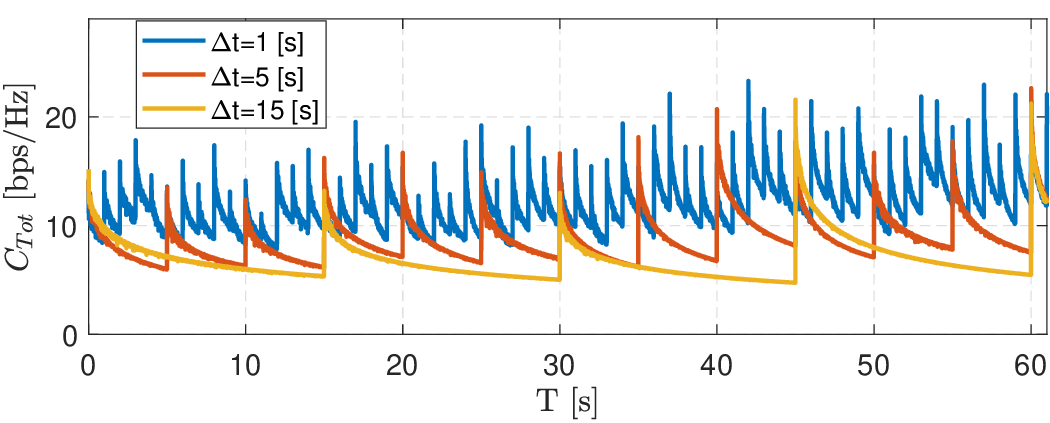}
\caption{Digital Beamforming \(M\times N=16\): Total system capacity performance with dynamic REM vs LEO orbiting time for \(K=2\) and \(J=2\).}
\label{fig:Digital_BF_Frequently_Updated_REM}
\end{figure}

A dynamic REM scenario was created, featuring three distinct REM time steps with \(K=2\) and \(J=2\). As the satellite traversed its orbit, it updated beamforming weights every 10 ms, while REM updates were received at intervals of \(\Delta t=\)1, 5, and 15 seconds. Both analog and digital beamforming were employed, and the results are presented in Fig. \ref{fig:Analog_BF_Frequently_Updated_REM} and \ref{fig:Digital_BF_Frequently_Updated_REM}. These figures reveal that performance peaks occur when the cognitive system is updated by the REM. Subsequently, the efficacy of the interference mitigation algorithm gradually diminishes until the next REM update is available. As mentioned in Section \ref{Sec_IMF}, it is computationally intensive for the satellite to perform beamforming multiple times in quick succession. Therefore, the REM update time step can be synchronized with the system's quality requirements to optimize the volume of on-board calculations while ensuring transmission quality.

\begin{figure}
\vspace{1mm}
\centering
\includegraphics[width=\linewidth]{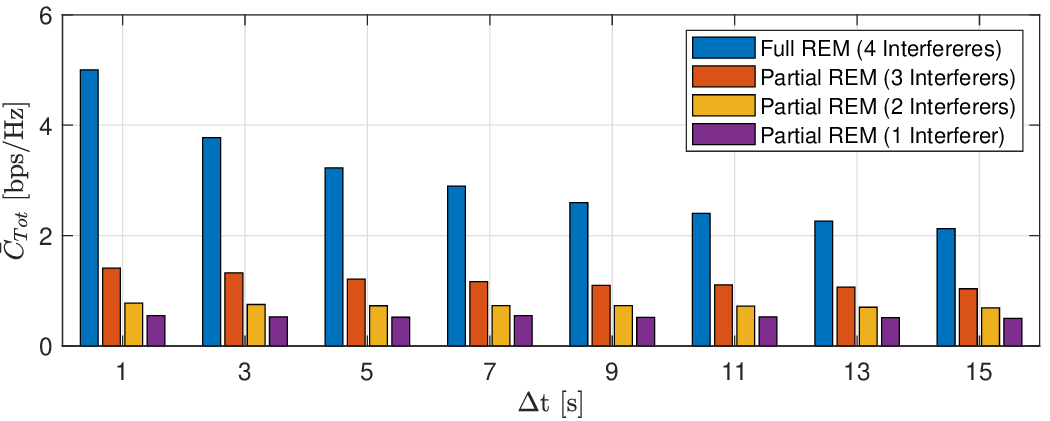}
\vspace{-7mm}
\caption{Analog Beamforming \(M\times N=16\): Mean total system capacity with partial vs full REM for \(K=2\) and \(J=4\).}
\label{fig:Analog_Full_Partial_REM}
\end{figure}

\begin{figure}
\centering
\includegraphics[width=\linewidth]{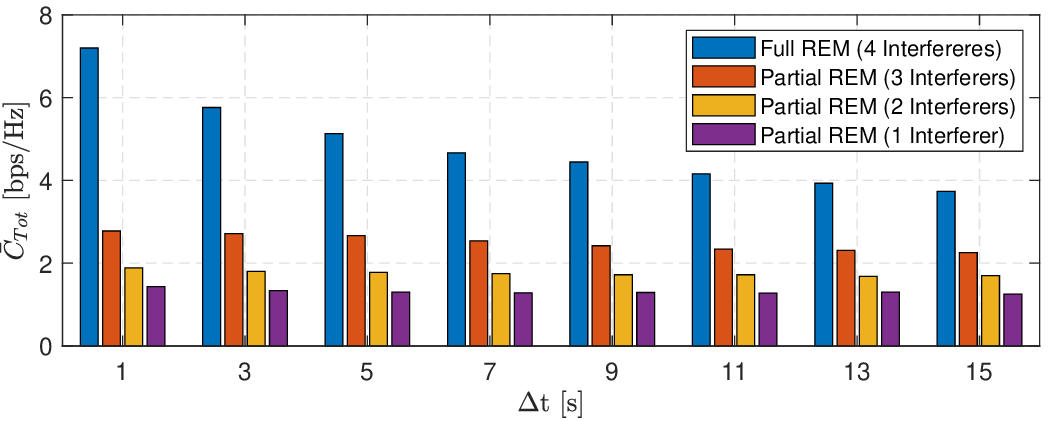}
\vspace{-7mm}
\caption{Digital Beamforming \(M\times N=16\): Mean total system capacity with partial vs full REM for \(K=2\) and \(J=4\).}
\label{fig:Digital_Full_Partial_REM}
\vspace{-5mm}
\end{figure}

We then examined a scenario comparing partial and full REM settings with \(K=2\) and \(J=4\), considering various REM update intervals for both analog and digital beamforming cases. Fig. \ref{fig:Analog_Full_Partial_REM} and \ref{fig:Digital_Full_Partial_REM} present the simulation results over a 150-second satellite movement. Blue bars represent the average total capacity when employing a \(M\times N=16\) planar array. In red, yellow, and purple, we observe the capacity performance achieved under partial REM conditions (\(Q=\)3, 2, and 1 interferers known to the REM).
These figures illustrate the impact of the REM update frequency on interference mitigation performance through beamforming, with this effect being more pronounced in the case of a full REM. Furthermore, there is a notable difference in the average total capacity between full REM and partial REM scenarios, underscoring the significance of cognitive capabilities within the system.

\begin{figure}
\centering
\includegraphics[width=\linewidth]{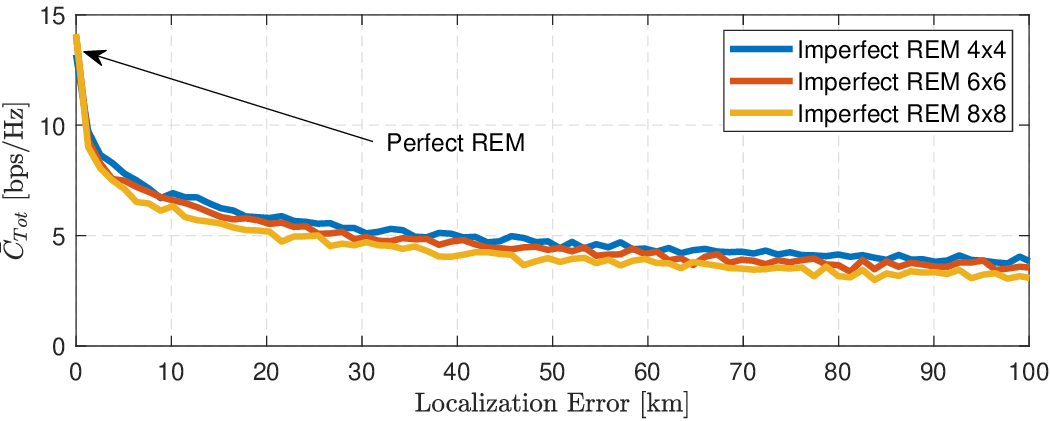}
\vspace{-7mm}
\caption{Digital Beamforming: Mean total system capacity under imperfect REM vs localization error.}
\label{fig:Digital_Error}
\end{figure}

We assessed an imperfect REM scenario involving a single interferer, for which the REM possessed location knowledge within a specified error range, as shown in Fig. \ref{fig:Digital_Error}. The location error spanned from 0 to 100 km. Three distinct antenna array sizes were employed to gauge the impact of location errors on interference mitigation. We used the suboptimal locations as input for the optimizer and calculated performance with the assumption that the interferer's position was perfectly known.
On the graph's leftmost side, we have the average capacity achievable when REM information about locations is flawless. As the location error grows, interference mitigation performance deteriorates, with a more significant impact observed when employing larger antenna arrays due to the enhanced resolution gained from increased antenna elements.
\vspace{-1mm}

\section{Conclusions}\label{Sec_Conclusions}
In this study, we conducted an analysis of interference mitigation in Low Earth Orbit (LEO) constellations with the assistance of limited Radio Environment Map (REM) information. We provided a comparative evaluation of various optimization algorithms for LEO beamforming and examined the performance of beamforming under three distinct REM information scenarios.
Our findings revealed that beamforming in LEO satellites is notably sensitive to the REM updating time steps, primarily due to the rapid variations in relative positions resulting from the orbital characteristics. Furthermore, we established that the comprehensiveness of information pertaining to interferers, as supplied by the REM, coupled with accurate position estimation, is of paramount importance in effectively mitigating interference.
\vspace{-2mm}
\section*{Acknowledgment}
We appreciate the support received from the Australian Government Research Training Program Scholarship, SmartSat CRC Scholarship, and the SmartSat CRC P1.27 Cognitive Satellite Radios Research Project.
\vspace{-2mm}
\bibliographystyle{IEEEtran}
\bibliography{IEEEabrv,Biblio}

\end{document}